# Effect of chromatic dispersion on image size and lattice plane curvature measurements with Rocking Curve Imaging

Elena Ferrari[a], Claudio Ferrari[a], Josè Baruchel[b], Thu Ni Tran Caliste[b], Stefano Basso[c], Akanshu Chauhan[c], Bianca Salmaso[c], Davide Sisana[c], Daniele Spiga[c]

[a]IMEM-CNR, Parco Area delle Scienze 37/A, 43124 Parma (Italy)
[b]European Synchrotron Radiation Facility (ESRF), 71 Avenue des Martyrs, F-38043 Grenoble (France)
[c]INAF Astronomical Observatory Brera, Via E. Bianchi 46, 23807 Merate, Lecco (Italy)

**Synopsis**   The effect of the difference between Si(111) double channel monochromator and sample Bragg angles in rocking curve imaging at synchrotron has been evaluated. This mismatch leads to a change in the image size parallel to the scattering plane which must be considered for accurate determination of resolution and lattice plane curvature.

**Abstract**   ADP crystals of large dimensions (80x80x20 mm$^3$), to be used as a X-ray beam expanders in the BEaTriX facility at INAF-OABrera, have been characterised at BM05 beamline at ESRF synchrotron with the main purpose to determine lattice plane curvature with a unprecedent accuracy, as the BEaTriX setup requires a radius of curvature larger than 22 km. In this beamline, the monochromator is made by 2 Si(111) parallel crystals in the non-dispersive configuration. Due to the difference in the Bragg angles between the Si(111) monochromator and the ADP(008) diffractions ($\vartheta_B^{Si}$=7.574° and $\vartheta_B^{ADP}$ =26.000° at E=15 keV respectively), only a limited part of the sample area, hit by the X-ray beam (11x11 mm$^2$ cross section), produced a diffracted beam for a given value of angle of incidence. In the rocking curve imaging techniques, a full image of the sample for a given peak position is obtained by combining images taken at different angles of incidence compensated by the dispersion correction, that is the Bragg condition difference in different points of the sample.

It is found that the resulting image size parallel to the scattering plane is affected by this dispersion correction. A 4.5 % elongation along the scattering plane was evaluated in the present Bragg case. This contribution is opposite in the Laue case.

# 1. Introduction

Rocking Curve Imaging (RCI) is a quantitative version of monochromatic beam diffraction topography that uses a two-dimensional detector, each pixel of which records its own 'local' rocking curve. Thanks to new 2D detectors with pixel resolution of a few microns, this technique permits the characterization of crystal defects and to locally measure the X-ray diffraction profile, the crystal bending, the strain and other crystal parameters (Tsoutsouva *et al.*, 2015; Lübbert *et al.*, 2000; Guo *et al.*, 2018; Caliste *et al.*, 2023; Mikulík *et al.*, 2003).

The crystal plane inclination map over the whole irradiated sample area can yield the overall crystal plane curvature, critical for high precision x-ray crystal optics. This technique was used to characterize the planarity of two 80x80x20 mm³ Ammonium Dihydrogen Phosphate (ADP) crystals, designed as beam expander for the 1.49 keV BEaTriX beamline. The facility is realized at INAF–Osservatorio Astronomico di Brera (Salmaso & Spiga) for the acceptance tests of the hundreds of modules composing the 2.5 m diameter optics of the NewAthena ESA mission. To ensure the beam collimation, the ADP crystal lattice must be defect free and perfectly flat (Radius of Curvatute R > 20 km) (Ferrari *et al.*, 2019).

As the detection of very small curvatures requires the measurement of very small angular shifts over large distances, systematic errors originated by the chromatic dispersion from the monochromator, must be corrected. Here we try to evaluate in detail such dispersion correction, both for reflection (Bragg) and transmission (Laue) geometry.

# 2. Calculations

In a typical RCI measurement the sample is placed at a distance R from the source and is irradiated by a beam of section $\Delta\vartheta_B^M \cdot R$, where $\Delta\vartheta_B^M$ is the divergence set by the slit aperture (Figure 1). The detector is located at a distance $d_1$ from the sample, with the surface perpendicular to the diffracted beam.
The beam diffracted by the monochromator, with Bragg angle $\vartheta_B^M$, has a wavelength dispersion $\Delta\lambda/\lambda$, in which each wavelength is diffracted at a different angular deviation $\Delta\vartheta^M$:

$$\frac{\Delta\lambda_M}{\lambda} = \frac{1}{\tan\vartheta_B^M} \cdot \Delta\vartheta^M \qquad (1)$$

where $\vartheta_B^M$ is the Bragg angle of Si(111) monochromator. The variation of the Bragg condition along the sample surface is an instrumental effect given by the wavelength dispersion, corresponding to an apparent crystal plane curvature. For any value of the angle of incidence $\Delta\vartheta^M$ on the sample surface the local Bragg angle deviation $\Delta\theta^S$ of the sample is given by:

$$\Delta\vartheta^S = \tan\vartheta_B^S \cdot \frac{\Delta\lambda_M}{\lambda} \quad (2)$$

So that the compensation $\Delta\vartheta^S - \Delta\vartheta^M$ of wavelength dispersion for any value of the angle $\Delta\vartheta^M$ is given by:

$$\Delta\vartheta^S - \Delta\vartheta^M = \left(\frac{\tan\vartheta_B^S}{\tan\vartheta_B^M} - 1\right) \cdot \Delta\vartheta^M \quad (3)$$

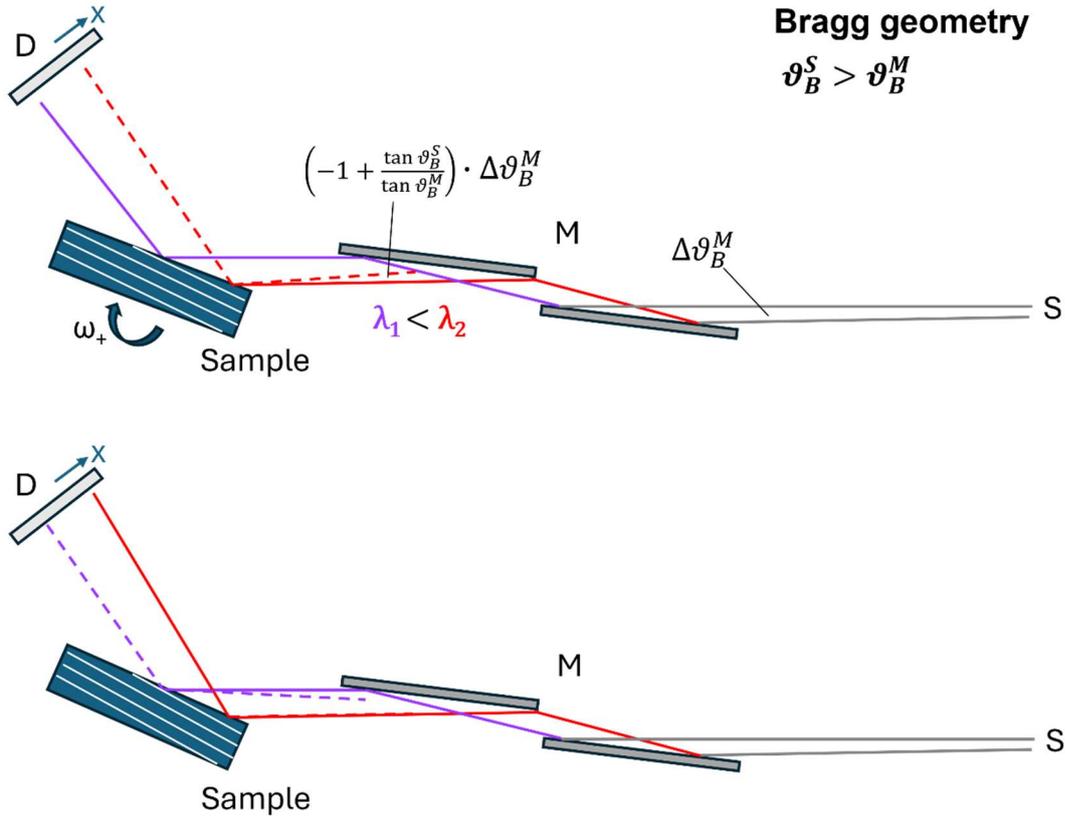

**Figure 1** Rocking curve imaging measurement in Bragg geometry with $\vartheta_B^S > \vartheta_B^M$, viewed perpendicular to the scattering plane. The upper and middle panel respectively shows the Bragg condition with $\lambda_1$ and $\lambda_2$, the two edge wavelengths of the beam. The dashed lines indicate the X-ray beams at Bragg condition for $\lambda_1$ and $\lambda_2$ wavelengths. Diffraction occurs only if the incidence angle matches the local Bragg condition ) angle during the goniometer rotation. These considerations hold for any angular deviation $\Delta\vartheta^M$ on the incident beam.

To calculate the dispersion correction $\Delta\vartheta^S(x) - \Delta\vartheta^M(x)$ at the pixel position x of the detector windows along the diffraction plane it is necessary to consider the cross section of incidence and diffracted beams. The effective angle $\Delta\vartheta^M(x)$ is obtained by considering the pixel position x, the distance detector-source R+d$_1$ and the ratio of the cross sections of the direct and the diffracted beam, called respectively $W_d$ and $W_s$. The resulting correction is:

$$\Delta\vartheta^S(x) - \Delta\vartheta^M(x) = \left(\frac{\tan\vartheta_B^S}{\tan\vartheta_B^M} - 1\right) \cdot \frac{x}{R} \cdot \frac{W_d}{W_s} \qquad (4)$$

The diffracted beam cross section $W_s$ is the sum of two contributions: one given by the direct beam size multiplied by the asymmetry factor $b$, the second by the variation due to divergence along the sample-detector path given by the monochromator and sample Bragg angle difference. The Bragg geometry is shown in Figure 2, upper panel. As the sample is rotated by $\left(\frac{\tan\vartheta_B^S}{\tan\vartheta_B^M} - 1\right) \cdot \Delta\vartheta^M$ to match the local Bragg condition, the diffracted beam rotates by $\left(2 \cdot \frac{\tan\vartheta_B^S}{\tan\vartheta_B^M} - 1\right) \cdot \Delta\vartheta^M$. Thus, the diffracted beam size $W_S$ at the detector position is:

$$W_s = R \cdot \Delta\vartheta^M \cdot |b| + \left(2 \cdot \frac{\tan\vartheta_B^S}{\tan\vartheta_B^M} - 1\right) \cdot \Delta\vartheta^M \cdot d_1 \qquad (5)$$

The second term in eq. 5 takes into account the diffracted beam size modification due to the sample to monochromator Bragg angle mismatch.

In the case of Laue geometry the positions of the two edge wavelengths of the beam are exchanged with respect to the Bragg geometry. Therefore, both the mismatch angle and the diffracted beam divergence have opposite sign than in Bragg case (Figure 2, middle panel).

To accurately evaluate the direct beam size $W_d$ in Eq. 4 we must first consider the detector pixel size measurement procedure. The apparent pixel size is measured by means of a calibrated pattern perpendicular to the beam at the sample position, with the detector at a distance $d_2$ (Figure 2, lower panel). Then the direct beam width is:

$$W_d = (R + d_2) \cdot \Delta\vartheta^M \qquad (6)$$

Substituting $W_d$ and $W_s$ in Eq. 4, with some calculations we finally get:

$$\Delta\vartheta^S(n_{px}) - \Delta\vartheta^M(n_{px}) = \pm\left(\frac{\tan\vartheta_B^S}{\tan\vartheta_B^M} - 1\right) \cdot \frac{n_{px} * l_{px}}{R} \cdot \frac{R + d_2}{R * |b| \pm \left(2 \cdot \frac{\tan\vartheta_B^S}{\tan\vartheta_B^M} - 1\right) \cdot d_1} \qquad (7)$$

Where $n_{px}$ is the pixel number and $l_{px}$ is the pixel size ( $n_{px} \cdot l_{px} = x$ pixel position); the signs +, - apply to Bragg and Laue geometries, respectively.

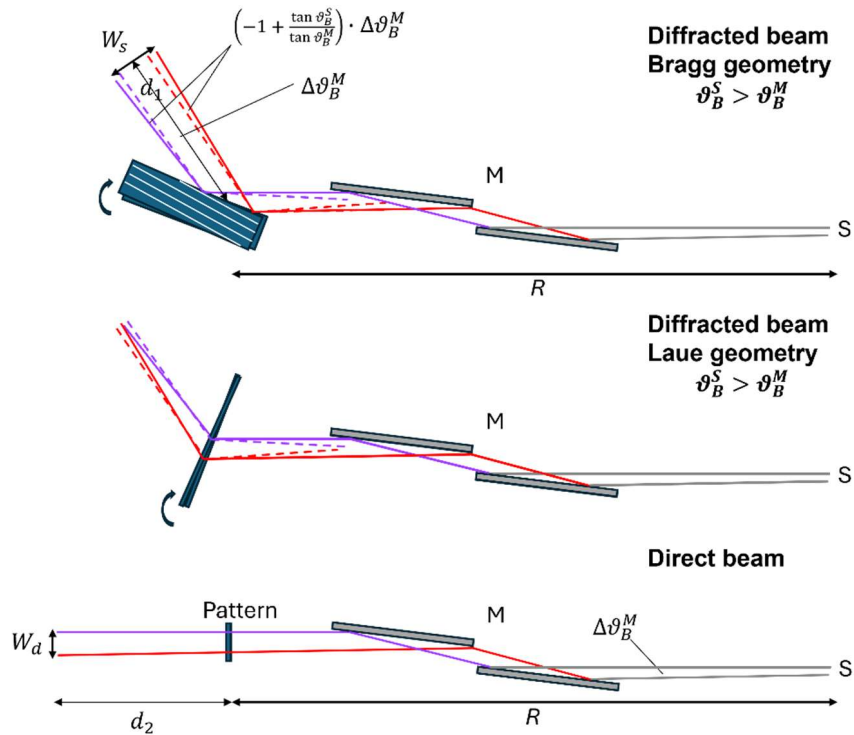

**Figure 2** Diffracted and direct beam considered in the calculation of the chromatic dispersion correction. The upper and the middle panel respectively show the Bragg and Laue geometries and represent the whole RCI acquisition. The overall diffracted beam divergence is twice the mismatch angle plus the direct beam divergence, but with opposite signs in the two geometries. The lower panel shows the pixel size measurement on the direct beam.

If the diffracted beam expansion along the sample-detector path is neglected, the $\frac{W_d}{W_s}$ ratio simplifies to $\frac{1}{|b|}$. However, such approximation leads to inaccurate chromatic dispersion correction and distance measurements on the image along the scattering plane.

The divergence effect was noticed during high accuracy lattice curvature measurements on two 80x80x20 mm³ Ammonium Dihydrogen Phosphate (ADP) crystals designed as beam expander for the 1.49 keV BEaTriX beamline at INAF–Osservatorio Astronomico di Brera (Spiga *et al.*, 2023). ADP has an orthorhombic cell with a = b = 7.53, c = 7.542 Å and is used in soft x-ray spectroscopy due to its strong (101) diffraction.

The RCI measurements were performed in Bragg geometry on ADP 008 planes with X-ray energy 15 keV at ESRF BM05 beamline equipped with a double Si(111) crystal monochromator. Here we consider only the ADP n°2 crystal, whose optical surface is inclined by (5.39±0.01)° and (0.13±0.01)° respect to the 001 planes along the [100] and [010] directions. RCI was performed with the scattering plane parallel either to [100] or [010] to estimate the lattice curvature along both directions. Experimental parameters relevant for Eq (6) are listed below:

- $\vartheta_B^M = 7.574°$    Bragg angle of Si(111) monochromator
- $\vartheta_B^S = 26.000°$   Bragg angle of ADP(008) sample
- $\Delta\vartheta_B^M = 37.2\ arcsec$   direct beam divergence
- $b = 1.48\ or\ 1.01$   asymmetry factors for 100 and 010 sample orientations
- $R = 61\ m$    Sample-Source distance
- $d_1 = 0.48\ m$    Sample-Detector distance for the RCI measurement (diffracted beam)
- $d_2 = 0.3\ m$    Sample-Detector distance for the pixel size measurement (direct beam)
- $l_{px} = 4.2\ \mu m$   pixel size

Fig 3 shows the RCI integrated intensity maps of the same part of ADP n° 2 crystal imaged with the two geometries as a function of pixel position on the detector. As the two images are deformed differently, the common area between them is identified with a red parallelogram. This area includes few dislocations.

We first evaluate the divergence effect on the image size comparing the pixel distances between dislocation pairs aligned along [010], measured with the two geometries. The selected pairs are named A-B and C-D in the maps. The [010] direction is respectively perpendicular and parallel to the scattering plane with the two geometries, thus corresponding to x and y in the two images. As the divergence effect occurs along x only, we compare the A-B and C-D distances measured along y or along x.

To improve the position accuracy we select the points of maximum intensity in the image, corresponding to the dislocation emergence points on the crystal surface. The calculated A-B and C-D distances are reported in Table 1.

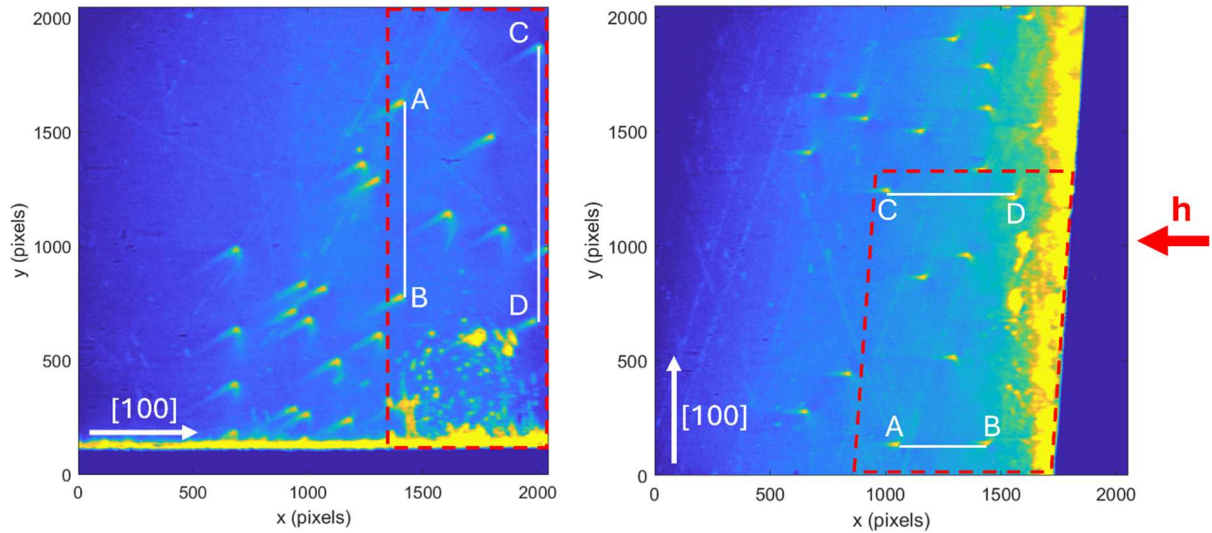

**Figure 3** RCI integrated intensity map of the same part of ADP n° 2 crystal, observed with the scattering plane parallel (left panel) or perpendicular (right panel) to the [100] direction. The common area on the two maps is marked with the red parallelogram and the letters indicate the dislocations chosen for the distance measurements. The incidence angle and surface tilt are respectively 20.61° and 0° in the left map and 25.87° and 5.4° in the right one.

**Table 1** Distances* calculated between the A-B and C-D dislocation pairs, measured perpendicular and parallel to the scattering plane, respectively corresponding to the left and right maps of Figure 3. The two distances are calculated as $\Delta y(px) \cdot l_{px}$ and $\Delta x(px) \cdot \frac{l_{px}}{\sin \theta_S} \cdot \frac{W_d}{W_s}$, with $\frac{W_d}{W_s}$ calculated neglecting the divergence or including it according to Eq 7. The exit angle $\theta_S$ is 26.13°.

| Defect pair | Distance (mm), measured | Distance(mm), divergence neglected | Distance (mm), divergence considered |
|---|---|---|---|
| A-B | 3.58±0.02 | 3.75±0.05 | 3.59±0.05 |
| C-D | 5.05±0.02 | 5.27±0.05 | 5.05±0.05 |

* $\Delta y(px)$ and $\Delta x(px)$ are obtained from the defect pixel coordinates in the maps of Fig 3: A (1401;1627), B (1399;774), C (2009;1872), D (1979;670) in the left map and A (1047;135), B (1440;139), C (1005;1244), D (1558;1216) in the right one.

If the beam expansion along the sample-detector path due to Bragg angle mismatch is neglected the distances along the scattering plane are overestimated by 4.5% . The resulting inaccuracy in the chromatic dispersion correction also affects the crystal lattice curvature estimation.

As for the Radius of Curvature measurement this is given by $RoC = \Delta x/\Delta\theta(x)$ , where $\Delta\theta(x)$ is the Bragg peak shift over the distance $\Delta x$ along the scattering plane. The angle $\Delta\theta(x)$ indicates the real lattice plane bending and is obtained by subtracting the chromatic dispersion $\Delta\vartheta^S(x) - \Delta\vartheta^M(x)$ (Eq. 4) from the measured Bragg peak shift $\Delta\omega(x)$:

$$\Delta\theta(x) = \Delta\omega(x) - \left(\Delta\vartheta^S(x) - \Delta\vartheta^M(x)\right) \qquad (8)$$

Figure 4 shows the peak position shift $\Delta\theta(x)$ of the left map in Figure 3. The distances are calculated neglecting or including the divergence, according to Eq. 7

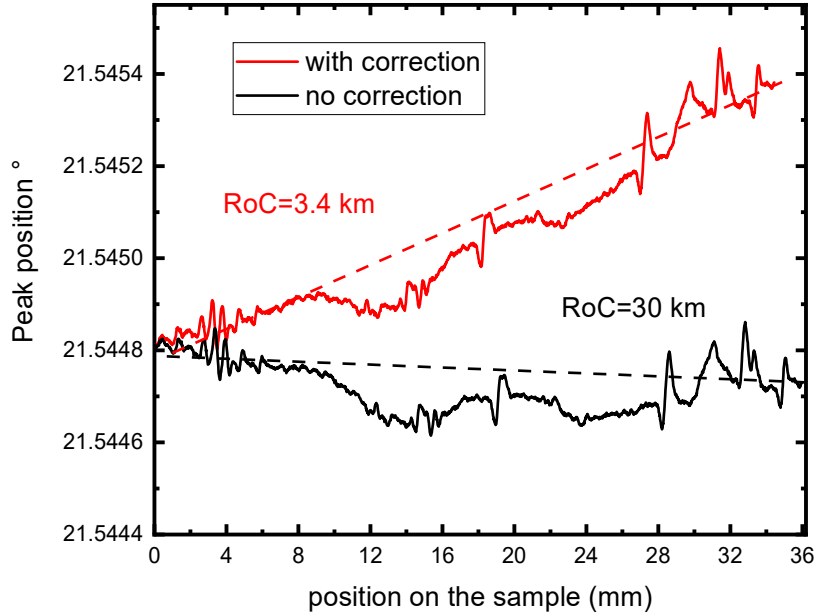

**Figure 4**: Bragg peak shift due to lattice plane curvature considering the beam divergence given by Bragg angle mismatch (red curve) or neglecting it (black curve) corresponding to RoCs of 3.4 and 30 km respectively.

A large discrepancy between the values of the Radius of Curvatures obtained with or without correction is apparent. Also the x sizes of the sample imaged by the detector differ by 4.5%, the shorter corresponding to the geometry corrected by the divergence expansion.

In conclusion we have found that for accurate measurements of distances and lattice plane curvatures in Rocking Curve Imaging technique it is necessary to consider the diffracted beam size change due to the difference between monochromator and sample Bragg angles in the direction parallel to the scattering plane. This effect corresponds to a reduction of image size for sample Bragg angle larger than monochromator Bragg angle for the Bragg case, and an enlargement in the Laue transmission case.

**Acknowledgements** E. F. acknowledges financial support from PNRR MUR project ECS_00000033_ECOSISTER. The BEaTriX team acknowledge financial support from ESA contract # 4000123152/18/NL/BW. The authors declare that there are no conflicts of interest.

**Data availability:** Data supporting the results obtained are reported in the manuscript.